\documentclass[journal]{IEEEtran}
\usepackage{blindtext}
\usepackage{graphicx}
\usepackage{amsmath}
\usepackage{amssymb}
\usepackage[noadjust]{cite}
\usepackage[mathscr]{euscript}
\usepackage{eufrak}
\usepackage{upgreek}

\usepackage{mdframed}

\usepackage{color}

\usepackage{color,soul}

\usepackage{algorithm,algorithmic}
\usepackage{framed}

\usepackage{lipsum}

\ifCLASSOPTIONcompsoc
\usepackage[caption=false, font=normalsize, labelfont=sf, textfont=sf]{subfig}
\else
\usepackage[caption=false, font=footnotesize]{subfig}
\fi

\newtheorem{remark}{Remark}
\newtheorem{definition}{Definition}

\newtheorem{proposition}{Proposition}

\begin{document}
	
	\title{Adaptive Tube-based Nonlinear MPC for Ecological Autonomous Cruise Control of Plug-in Hybrid Electric Vehicles }
	
	\author{Bijan Sakhdari, Nasser L. Azad}

	\maketitle

	\begin{abstract}
			This paper proposes an adaptive tube-based nonlinear model predictive control (AT-NMPC) approach to the design of autonomous cruise control (ACC) systems. The proposed method utilizes two separate models to define the constrained receding horizon optimal control problem. A fixed nominal model is used to handle the problem constraints based on a robust tube-based approach. A separate adaptive model is used to define the objective function, which utilizes least square online parameter estimators for adaption. By having two separate models, this method takes into account uncertainties, modeling errors and delayed data in the design of the controller and guaranties robust constraint handling, while adapting to them to improve control performance. Furthermore, to be able implement the designed AT-NMPC in real-time, a Newton/GMRES fast solver is employed to solve the optimization problem. Simulations performed on a high-fidelity model of the baseline vehicle, the Toyota plug-in Prius, which is a plug-in hybrid electric vehicle (PHEV), show that the proposed controller is able to handle the defined constraints in the presence of uncertainty, while improving the energy cost of the trip. Moreover, the result of the hardware-in-loop experiment demonstrates the performance of the proposed controller in real time application.
	\end{abstract}

	
	\section{Introduction}
	\IEEEPARstart{T}{he} current number of vehicles in the world is approximately 1 billion, a number that – with the existing high demand for personal transportation, – is expected to double over the next few decades. \cite{PaulADAS}. The rapid growth in the number of vehicles has resulted in high air pollution, increased energy demand, higher traffic intensity, longer travel times and increased risk of accidents. In fact, the annual cost of traffic injuries worldwide is estimated at \$518 billion \cite{piao2008advanced}. Approximately 1.3 million people die in car accidents each year; this number is expected to increase to 1.9 million annually, based on the current growth rate in vehicle ownership  \cite{alam2014fuel}.  At the same time, air pollution caused by the transportation sector is escalating in many parts of the world.
Greenhouse gas emissions produced by the transportation
sector has doubled in the past few decades, and is currently responsible for approximately 22\% of the total anthropogenic global greenhouse gas emissions[4]. Issues such as these have provided the basis for many researchers to pursue the development of safe and efficient transportation systems using modern technology. 
	
	In the recent years, advancements in embedded digital computing and communication networks have enabled the development of automated driving systems, with Advance Driver Assistance Systems (ADAS) being one of the most important of these developments. ADAS technologies assist drivers according to the changes in the environment based on sensing the vehicle’s surroundings \cite{PaulADAS}. ADAS can reduce the effect of error in human judgment in emergency situations to improve driving safety and performance by making optimal decisions to enhance the autonomous interaction with the environment. Autonomous Cruise Control (ACC) is a topic of interest among several types of ADAS, spurring many scientific studies in the field. ACC is an advanced version of the cruise control system that has the ability to automatically accelerate or decelerate, without any additional input from the driver, when the preceding vehicle is speeding up or slowing down. This system is beneficial in many ways, as it can improve traffic flow, reduce the possibility of accidents and provide a comfortable driving setting through a semi-autonomous driving experience. \cite{IoannouChien}.
	
	In order to maintain a safe inter-vehicular distance, the most commonly developed ACC system utilizes linear controllers. Authors of \cite{[4]}, in their study of adaptive cruise control-based concept, developed a single-lane ACC with intelligent ramp metering by enabling vehicles to move in short inter-vehicular distances to increase highway capacity. In \cite{[5]}, a more complex ACC system was created with the ability to adapt itself to differing driving/road conditions. The researchers applied an algorithm to automatically detect traffic conditions based on surrounding information, which then subsequently changed the parameters of the ACC system in accordance with each traffic situation. In \cite{[6]}, authors developed a PID-based ACC aiming to perform and behave as a human driver would. Other authors employed linear methods in their design \cite{[7],[9]}, resulting in low-complexity tracking performance. These methods could provide a satisfactory tracking performance with low complexity. Also, PID controllers offer tuning parameters that can be easily adjusted to match different situations and systems. 
	
	To improve the benefits of ACC systems, it is possible
to consider fuel efficiency in its design. Considering future prediction of traffic motion and environment of the vehicle could be very useful in this direction  \cite{[10], sakhdari2016ecological}. Utilizing ACC to improve fuel efficiency of the vehicles has been widely investigated in previous research. In \cite{wang2012driver}, the authors developed an Ecological ACC (Eco-ACC) system based on Model Predictive Control. They assumed a stationary condition (zero acceleration) on surrounding vehicles in order to perform their finite horizon optimization. Their result shows that utilizing future prediction of the preceding vehicle’s trajectory yields better fuel economy. In  \cite{[10]}, a smooth acceleration degradation in the prediction horizon was employed to predict the preceding vehicle’s trajectory and a jamming wave prediction was presented to prevent jamming waves while maintaining a safe inter-vehicular distance. The authors used MPC to address this problem and showed that their method improves traffic flow and driver comfort, as well as fuel efficiency. In  \cite{vajedi2016ecological},\cite{sakhdari2016ecological} a higher energy efficiency has been achieved via exploiting nonlinear MPC in designing Eco-ACC for Plug-in Hybrid Electric Vehicles (PHEV).
	
	The majority of studies in this area to date assume that radar measurements are reliable with little to no imperfection or uncertainty. However, this assumption is not realistic as radar and Lidar performance are highly dependent on weather conditions and weather elements, such as snow, rain, or fog, which can substantially reduce their accuracy  \cite{[12], [13]}. The sensors’ precision is also contingent on the number of objects within range and the type of background \cite{[14]}. In order to maintain its stability and performance, a reliable ACC must be resilient to data uncertainty and modeling error. With wind and road disturbances as their main consideration, the authors in \cite{[15]}, developed an $H_2/H_\infty$ control method-based model-based controller, which lead to improved tracking performance demonstrated in simulations. In \cite{[16]}, the authors assumed a linear model for their vehicle with disturbances on the states to develop a robust ACC system. This led to improved tracking performance, but also resulted in higher computational effort through the utilization of min-max robust MPC method, which makes this method not suitable for real-time applications. Another version of robust MPC is tube-based MPC (T-MPC), which works based on a tube resulted from bounded uncertainties in the system \cite{[17], [18]}. In order to ensure the tight bondage of the real states inside the demarcated restraints, T-MPC maintains the nominal system inside a tighter area. The computational demand of T-MPC is only slightly higher than regular MPC, since the requisite tube can be calculated offline, making it suitable for real-time applications.  In \cite{gray2013robust} and \cite{gao2014tube}, the authors used T-MPC to design a semi-autonomous ground vehicle. They considered the uncertainties and nonlinearities as an additive disturbance, then calculated a tube for the disturbed states. Their MPC used the resulted tube to gain robustness against system uncertainties. In \cite{myITSc}, the authors developed a robust ACC controller using linear T-MPC. Their simulation on a high-fidelity vehicle model showed that this method can ensure robustness against delayed data, uncertainty and modeling errors in a car following scenario.
	
	Robust control methods can guarantee safety and stability; however, they are usually conservative and can deteriorate performance of the controlled system. Therefore, many researchers prefer adaptive control approaches that estimate changes in parameters and respectively adapt to them to maintain performance and stability of the system. Moreover, due to changing conditions and aging of the vehicle, the actual parameters of a car might change and, therefore, an on-line optimization algorithm with a fixed model might not be able to find the actual optimal control decisions. To consider this matter, in \cite{zhang2017hierarchical}, the authors designed a hierarchical cruise control for connected vehicles and used a gradient-based parameter estimation to estimate the changes in vehicle parameters. They fed the estimated parameters to a low-level sliding-mode controller to regulate axle torque so that the desired states are followed. In \cite{santin2016cruise}, authors used a recursive least square parameter estimator and adaptive nonlinear MPC to design a cruise controller with fuel optimization. They used parameter estimation to improve the control-oriented model of their MPC and, by performing vehicle experiments, showed that their method can achieve a 2.4\% improvement in fuel economy, compared to a production cruise controller. Similar adaptive control approaches can be found in \cite{kwon2014adaptive},\cite{lin2007car},\cite{lee2003adaptive},\cite{SwaroopHedrick}. Although these methods can capture the changes in the model and act accordingly, in the event of a sudden change in parameters or wrong estimation, they may loose performance and stability. Specially, for close car following, the controller must be able to guarantee safety of the system while improving the performance. Therefore, a method is needed that can adapt to changes while being robust to uncertainties, disturbances and model errors. To combine robustness and model adaptation, in \cite{aswani2013provably} and \cite{aswani2012extensions} authors used a type of adaptive MPC that they called learning-based model predictive control. In their method, a linear controller generates optimal inputs based on a learned linear model, while a separate model checks if the constraints will be satisfied. Nonlinear learning based MPC was used in  \cite{ostafew2014learning} and \cite{ostafew2016learning} for path tracking control of a mobile robot in outdoor and off-road environment. They used a simple known model and a Gaussian process disturbance model that can be learned based on trial experience. Their experimental results on different robot platforms show that their controller is able to reduce path tracking error by learning and improving disturbance model through experience. 
	 
	In this paper, an adaptive tube-based nonlinear MPC (ATNMPC) controller is presented that can improve the performance of Eco-ACC by adapting to the changes in system, while maintaining robustness against uncertainties, disturbances and modeling errors. This method decouples performance from robustness and, therefore, is able to maintain stability and safety while adapting to changes in the system and the environment. First, nonlinear T-MPC method is used to design a robust controller that can handle the uncertainties in estimation of the drag coefficient, gravitational forces due to uncertainty in road’s grade estimation, uncertainty in the preceding vehicle’s acceleration, and also delay in the data gathered from the on-board vehicle radar. The designed controllers optimizes vehicle’s motion in finite horizon to improve the consumed energy cost of the vehicle, while handling the defined constraints in the pretense of uncertainty and disturbances. Then, to capture changes in the system and enhancing the control-oriented model, an on-line parameter estimation algorithm is used that estimates new parameter values based on minimizing the error between estimated and actual output of the system. This way the on-line optimization will find the actual optimal point based on the adapted model while constraints are handles based on the original nominal model.

	The main contribution of this paper is in combining robustness against uncertainties with parameter adaptation in the design of controller. The TA-NMPC approach is proposed that can robustly handle defined constraints while adapting to the changes and improving performance of the Eco-ACC system for PHEVs. Moreover, to be able to execute the designed optimal controller in real-time, a Newton/GMRES fast solver is adapted to solve the AT-NMPC optimal control problem. The rest of the paper is structured as follows: Section II presents preliminary definitions; In section III, modeling procedure is explained and a model for car-following is presented that can be used for the design of ACC controllers, also uncertainty bounds are defined based on the presented model as an additive disturbance. In section IV, the controller design is explained by taking advantage of the AT-NMPC method to achieve robustness and high performance in ACC design. Section V is dedicated to control evaluation. In this section, the proposed method has been simulated on a high-fidelity vehicle model in a car-following scenario and in a simulation environment with injected uncertainties. Moreover, HIL experiments has been presented in this section for the proposed controller.  Finally, conclusions are presented in Section VI.

	\section{PRELIMINARIES}
	In this paper, $\oplus$ and $\ominus$  indicate the Minkowski sum and the Pontryagin's set difference. If $X$ and $Y$ are sets, then: $Y\oplus X=\{x+y:x\in X,y\in Y\}$ and $Y\ominus X=\{v\in R^n:v\oplus X\subseteq Y\}$. Also, $X\oplus (Y\oplus V)=\{x+y+v:x\in X,y\in Y,v\in V\},AX=\{Ax:x\in X\},AX+BY=\{Ax+By:x\in X,y\in Y\}$  and $\oplus^{m}_{i=n} Y_i=Y_n+Y_{n+1}+\dots +Y_m$.
	
	\section{Modeling}
	This section explains the models that have been used for design and evaluation of the proposed controller. Control evaluation has been done on a high-fidelity model of the base-line vehicle, which is a Toyota Plug-in Prius. It consists of complex high-fidelity models and mappings of all the components in the vehicle that can affect longitudinal motion and energy consumption. The high accuracy of this model makes it a reliable tool for evaluation of the designed controllers. For control design, however, a simple model is needed that has low computational demand but is descriptive enough to capture the general behavior of the system. Here, different control-oriented models are presented that represent longitudinal motion and energy consumption of the vehicle.
	\begin{figure}
		\includegraphics[scale=0.39]{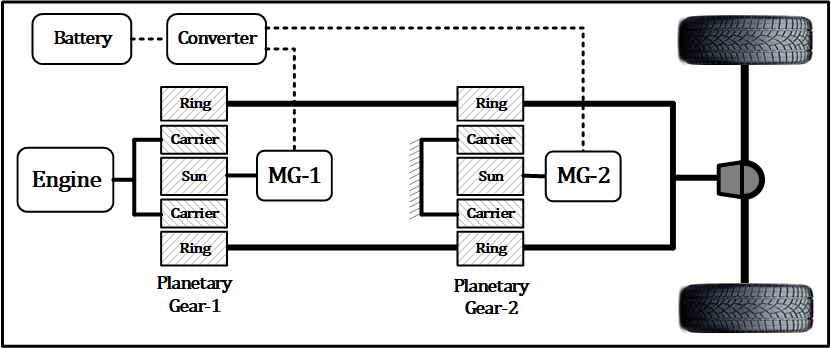}
		\centering
		\makeatletter
		\caption{Schematic of Toyota power-split power-train}
		\label{fig:powertrain}
	\end{figure}

	\subsection{High-fidelity power-train model}
	Fig. \ref{fig:powertrain} illustrates the power-train of a power-split Toyota Plug-in Prius. It has an internal combustion engine and two electric motors that combined with each other and through planetary gears, power the vehicle. To evaluate performance of the designed controllers, a high-fidelity model of the vehicle is needed. Therefore, a model of the base-line vehicle was developed in Autonomie which is a new generation of PSAT software developed by Argon National Lab. Autonomie has a library of vehicle models and components that can be selected to generate a high-fidelity model of the whole vehicle. It allows the selection of a two wheel vehicle with hybrid power-split power-train and modification of its components and characteristics to simulate the base-line PHEV. The procedure taken in development of the high-fidelity model and testing its validity has been presented in previous publications by the author’s research group  \cite{taghavipour2013high},\cite{azad2016chaos}. This model has been used for evaluation of the proposed controllers in terms of longitudinal motion control and trip energy cost.
		
	\subsection{Car following control-oriented model}
\begin{figure}
		\includegraphics[scale=0.083]{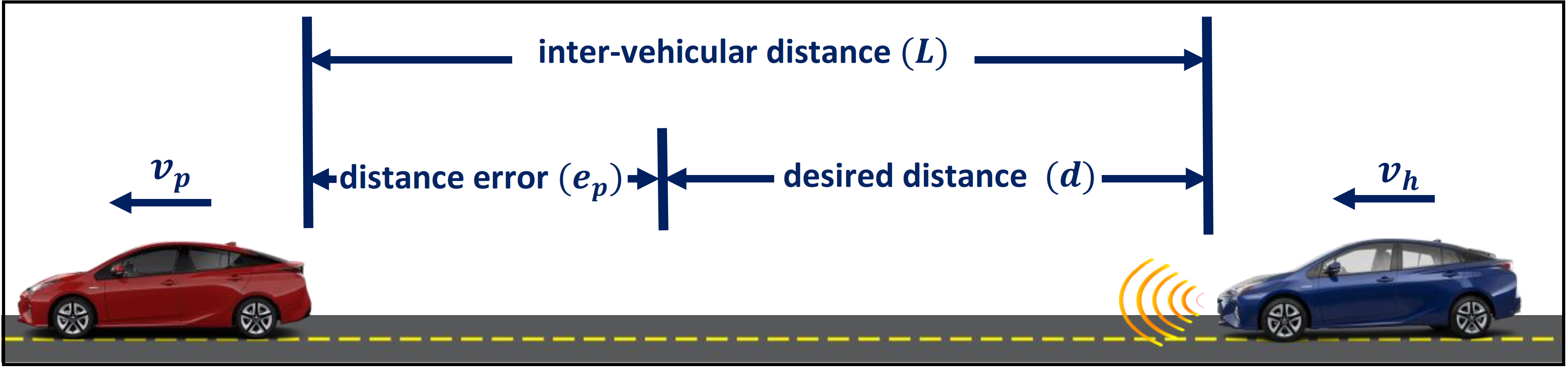}
		\centering
		\makeatletter
		\caption{Car-following with autonomous cruise control}
		\label{fig:carfollowing}
	\end{figure}

	To develop a model for car-following problem shown in the Fig. \ref{fig:carfollowing}, it is necessary to define a safe car following rule. Among different spacing policies in literature, constant time headway rule was chosen in this paper: $d=d_0+hv_h$, where $d$ is the desired distance, $d_0$ is the minimum distance at stand still, $v_h$ is the host vehicle's velocity and $h$ is the constant headway time. This spacing policy requires increasing distance with respect to velocity so it takes a specific constant amount of time for the host vehicle to reach to its preceding. Based on the chosen gap policy, the sate equations of the system can be written as follows:
	\begin{equation}\label{state_equation}
	\setlength{\jot}{6pt}
	\begin{aligned}
	\begin{split}
	&\dot{x} = A x + B u(t-\tau_a) + B_p a_p + B_g F_r \\
	&x= \begin{bmatrix} e_p\\e_v\\v_h\\T_w \end{bmatrix}, \quad A= \begin{bmatrix}
		0 & 0 & 1 & -\frac{h}{m r_w} \\ 0 &0 &0 &-\frac{1}{m r_w} \\ 0 &0 &0 &\frac{1}{m r_w} \\ 0 &0 &0 &-\frac{1}{\eta} \end{bmatrix}, \\
	&B= \begin{bmatrix}
	0 \\ 0 \\ 0 \\ \frac{K_a}{\eta}
	\end{bmatrix}, \quad 
	B_p = \begin{bmatrix} 
	0 \\ 1 \\ 0 \\ 0 
	\end{bmatrix}, \quad
	B_g= \begin{bmatrix}
	-h \\ -1 \\ 1 \\ 0
	\end{bmatrix}, \\ 
	&F_r = -\frac{1}{2} \rho_a A_c C_d (v_h+v_w )^2-(\mu_{r_0}+\mu_{r_v} v_h) mg\cos(\phi_r) \\
	& \quad  -mg\sin(\phi_r ), \\
	\end{split}
	\end{aligned}
	\end{equation}
	`
	To define a cost function based on energy economy improvement, a control-oriented model for the consumed energy is needed. Because our base-line vehicle is a PHEV, we have to consider both fuel and electricity costs. Therefore, instead of fuel rate and electrical current, we define the cost function based on combined energy cost of the two sources
	
	\begin{equation}\label{eq3}
		\begin{aligned}
			\begin{split}
				E_{cost} = -C_f \frac{ \dot{m}_{f}}{v_h}- C_e \frac{\dot{SOC}}{v_h},
			\end{split} 
		\end{aligned}
	\end{equation}
	where $E_{cost}$ is the cost of energy, $C_f$ and $C_e$ are the cost of gasoline and electricity, respectively, and $SOC$ is the state of charge of the battery. The energy cost has been divided by the velocity to eliminate the effect of traveled distance. At lower velocities, the energy cost will be assumed constant to avoid singularities. 
	An energy management algorithm decides the distribution of energy between the energy sources while the vehicle is running to keep the power-train near its optimal working point. Therefore, it can be assumed that the engine is always working in its optimum working point and approximate fuel consumption with the following equation \cite{8013751}:
	\begin{equation}\label{eq4}
		\begin{aligned}
			\begin{split}
				\dot{m}_{f_i}=\alpha_1+\alpha_2 P_{e} + \alpha_3 P_{e}^2 + \alpha_4 v_h,
			\end{split} 
		\end{aligned}
	\end{equation}
	\color{black} {where $\dot{m}_{f_i}$ is the fuel rate, $P_{e_i}$ is the engine power and $\alpha_1$, $\alpha_2$, $\alpha_3$ and $\alpha_4$ are constant coefficients}\color{black}. To estimate the electricity rate, we used the following equation:
	
	\begin{equation}\label{eq5}
		\begin{aligned}
			\begin{split}
				\dot{SOC}=\gamma_1+\gamma_2 P_m + \gamma_3 P_m^2,
			\end{split} 
		\end{aligned}
	\end{equation}
	where $P_m$ is the motors' or generators' power and $\gamma_1$, $\gamma_2$ and $\gamma_3$ are constant coefficients. The squared electric power has been included in the model to represent ohmic losses. 
	As mentioned, energy management decides the power ratio between electricity and gasoline. Therefore, based on power ratio, the energy cost can alternate for different total power demands. Based on power ratio the power demand from each source can be calculated:  
	\begin{equation} \label{eq6}
		\begin{aligned}
		\begin{split}
				PR &= \frac{P_e}{P_{total}},\\
				P_e &=PR*v_h*u*m,\\
				P_m &=(1-PR)*v_h*u*m,\\				
		\end{split}
		\end{aligned}
	\end{equation}
	where $PR$ is power ratio and $P_{total}$ is the total power demand. 
		
	\subsection{Reduced model}
	The control-oriented model needs to be updated based on the on-line measurements in the system. In this paper, adaptation has been done based on a recursive Least-square method, which requires a parametric model. The following formulation has been used for longitudinal dynamic's parametric model:
	
	\begin{equation} \label{long_est_model}
		\begin{aligned}
			\begin{split}
				a_h= &\frac{R_{g} \eta_{p}}{r_{w} m} T_{com} - \frac{\rho_a A_c C_d}{2 m} (v_h+v_w)^2\\ &- g(\mu_{r_0}+\mu_{r_v})cos(\phi_r)-g sin(\phi_r),	\\
			\end{split}
		\end{aligned}
	\end{equation}
	where $R_g$ is gear ratio, $\eta_p$ is power-train efficiency, $T_{com}$ is the commanded torque and other parameters are as defined before. This formulation can be rearranged as follows:
	\begin{equation} \label{eq8}
		\begin{aligned}
			\begin{split}
				s v_h= &\frac{R_{g} \eta_{p}}{r_{w} m} T_{com} - \frac{\rho_a A_c C_d}{2 m} (v_h)^2\\ 
				& - (\frac{\rho_a A_c C_d}{m} v_w +g \mu_{r_v}) (v_h)-g(\phi_r) \\
				& 	- \frac{\rho_a A_c C_d}{2 m} {v_w}^2 -g\mu_{r_0}.	\\
			\end{split}
		\end{aligned}
	\end{equation}	
	Finally, by using stable filtering this model can be reduced to the following model:
	\begin{equation} \label{parametrimodel_long}
		\begin{aligned}
			\begin{split}
				\frac{s}{s+\lambda} v_h =  &\theta_1 \frac{T_{com}}{s+\lambda} -\theta_2 \frac{{v_{h}}^2}{s+\lambda}-\theta_3 \frac{v_{h}}{s+\lambda} \\
				&-\theta_4 \frac{\phi_{r}}{s+\lambda}-\theta_5 \frac{1}{s+\lambda},
			\end{split}
		\end{aligned}
	\end{equation}	
	where $\lambda$ is the stable filter's time constant. Therefore the reduced parametric model is:
	\begin{equation} \label{eq10}
		\begin{aligned}
			\begin{split}
				\hat{a}_h = \hat{\Theta}^T \Phi_d,
			\end{split}
		\end{aligned}
	\end{equation}		
	where $\hat{a}_h$ is the estimated acceleration and:
	\begin{equation} \label{eq10}
		\begin{aligned}
			\begin{split}
				\hat{\Theta}&= \begin{bmatrix} \hat{\theta}_1 & \hat{\theta}_2 & \hat{\theta}_3 & \hat{\theta}_4 & \hat{\theta}_5 	\end{bmatrix}^T ,\\
			\end{split}
		\end{aligned}
	\end{equation}
	which $\hat{\Theta}$ is the vector of the estimated parameters and:
	\begin{equation} \label{eq11}
		\begin{aligned}
			\begin{split}
				\Phi_d&=\frac{1}{s+\lambda}\begin{bmatrix} T_{com} &-{v_h}^2 &-v_h &\phi_r &1 \end{bmatrix}^T,
			\end{split}
		\end{aligned}
	\end{equation}	
	where $\Phi_d$ is the regressors' vector for longitudinal dynamic estimator. Equations \ref{eq4} and \ref{eq5} are already in the parametric form of $\dot{m}_f=\hat{A}\Phi_e$ and $\dot{SOC}=\hat{\Gamma}\Phi_m$ with:
	\begin{equation} \label{eq12}
		\begin{aligned}
			\begin{split}
				\Phi_e &=\begin{bmatrix} 1 &P_e &{P_e}^2 &v_h \end{bmatrix}^T,\\
				\Phi_m &=\begin{bmatrix} 1 &P_m &{P_m}^2  \end{bmatrix}^T,\\
				\hat{A} &=\begin{bmatrix} \hat{\alpha}_1 & \hat{\alpha}_2 & \hat{\alpha}_3 & \hat{\alpha}_4  \end{bmatrix}^T,\\
				\hat{\Gamma} &=\begin{bmatrix} \hat{\gamma}_1 & \hat{\gamma}_2 & \hat{\gamma}_3  \end{bmatrix}^T.
			\end{split}
		\end{aligned}
	\end{equation}		
	The hat shows the estimated value of a parameter. Presented models in this section will be used for control design and evaluation in the following sections. 
	
	\section{Control design}
		This section is devoted to the control design procedure. First, the effective disturbances and uncertainties are analyzed and an additive disturbance term that captures them is presented. A linear feedback controller (Kc) is designed that stabilizes the system and bounds the effect of additive disturbances on the system's sates. Then, nonlinear T-MPC design procedure is explained, which is able to handle the defined constraints in the presence of bounded uncertainties and disturbances. The final control input to the system is generated by combining the designed linear controller with the output of MPC. Finally, an on-line least square parameter estimator is presented,  which estimates the uncertain parameters of the system in real-time. The estimated parameters are used inside the MPC controller to improve its performance in case of a change in the parameter values. This way the final system will be robust to changes in the uncertain parameters that can also adapt to them to improve the control performance. Therefore, robustness and performance will be decoupled and achieved simultaneously. Fig. 3 illustrates the proposed Eco-ACC architecture.
		
		\begin{figure}
			\includegraphics[scale=0.36]{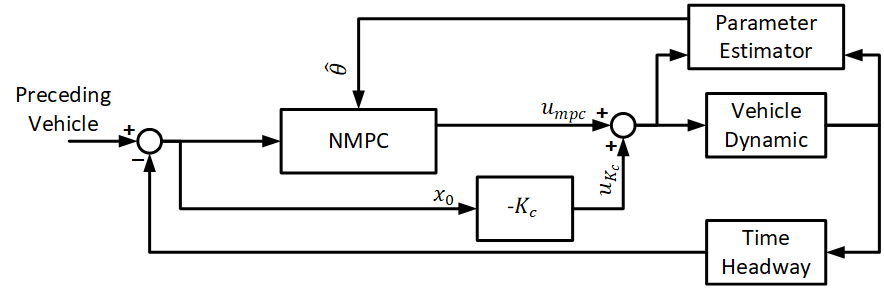}
			\centering
			\makeatletter
			\caption{AT-NMPC controller architecture}
			\label{fig:mesh6}
		\end{figure}
		
		\subsection{Disturbance set}
		
		To design a robust MPC, a bound must be established on the states' error caused by disturbances or a robust positive invariant set defined below.
		
		\begin{definition}
			For an autonomous system $x[k+1]=Ax[k]+Bw[k]$ with bounded disturbance $w[k] \in \mathbb{W}$, robust positive invariant set $\Phi$ is the set of all $x[k]\in\Phi$ such that for all $w[k] \in \mathbb{W}$ and $i>0$, $x[k+i]\in\Phi$ \cite{[20]}. 
		\end{definition}
	
		Suppose a nonlinear system in the following format:
		
		\begin{equation} \label{disturbed_model}
			\begin{aligned}
				\begin{split}
					&x[k+1]=Ax[k]+Bu[k]+g(x[k])+w[k],\\
					&\text{subject to}\\
					& \quad \quad x[k] \in \mathbb{X}, \\
					&  \quad \quad u[k] \in \mathbb{U},\\
					& \quad \quad w_[k] \in \mathbb{W}_m,\\
				\end{split}
			\end{aligned}
		\end{equation}			
		where $x$ is the state of the system, $u$ is input, $w$ is an additive disturbance, $\mathbb{X}$, $\mathbb{U}$ and $\mathbb{W}_m$ are bounds on state, input and disturbance in the system and $g(x)$ is the nonlinear part of the system.
		Now suppose that the input to the system has the following form:
		\begin{equation} \label{eq13}
			\begin{aligned}
				\begin{split}
					u_k = -K x[k-\uptau_d]+ c_0 
				\end{split}
			\end{aligned}
		\end{equation}	
	where $K$ is a linear stabilizing controller, $c_0$ is the input generated by the model predictive controller and $\uptau_d$ is the total delay in radar and actuation. We ignore $\uptau_d$ in the rest of calculations and model it as part of uncertainty in Proposition 1. By considering this input, (\ref{disturbed_model}) can be rewritten as
		\begin{equation} \label{disturbed_model_stbl}
			\begin{aligned}
				\begin{split}
					x[k+1]=A_c x[k]+B c_0 +g(x[k])+w[k],
				\end{split}
			\end{aligned}
		\end{equation}	
	where $A_c = A-BK$. On the other hand, without considering the disturbance term, the nominal system can be written as
		\begin{equation} \label{nominal_model}
			\begin{aligned}
				\begin{split}
					\bar{x}[k+1]=A_c \bar{x}[k]+B c_0 +g(\bar{x}[k]),
				\end{split}
			\end{aligned}
		\end{equation}
	where $\bar{x}$ is the nominal state. By reducing (\ref{nominal_model}) from (\ref{disturbed_model_stbl}) and considering state error as $e=x-\bar{x}$, error dynamic can be defined.
	\begin{equation} \label{error_model}
	\begin{aligned}
	\begin{split}
		e[k+1] = A_c e_k + (g(x[k])-g(\bar{x}[k])) + w[k]
	\end{split}
	\end{aligned}
	\end{equation}
	An RPI set of this system is equivalent to the maximum error caused by the additive disturbance. To be able to find RPI set of this system, we need to handle the error in the nonlinear term. Authors of \cite{gao2014tube} showed that if the nonlinear term $g(x)$ is Lipschitz continues, $\left\Vert g(x)-g(\bar{x}) \right\Vert_2$ can be bounded. If $g(x)$ is Lipschitz in the region $x\in \mathbb{X}$ then:
	\begin{equation}\label{Lipschitz_def}
	\left\Vert g(x)-g(\bar{x}) \right\Vert_2 \leq L \left\Vert x-\bar{x} \right\Vert_2, \quad \forall x_1,x_2 \in \mathbb{X},
	\end{equation}
	where the smallest $L$ satisfying this condition is the Lipschitz constant. Now if $L(\mathbb{X})$ is the Lipschitz constant over $\mathbb{X}$ then it can be obtained from (\ref{Lipschitz_def}) that
	 \begin{equation}
	 \begin{aligned}
	 \begin{split}
	 \forall x,\bar{x} \in \mathbb{X} \;\, \& \;\, e \in \mathbb{E}, \: \left\Vert g(x) -g(\bar{x}) \right\Vert_\infty \\ \leq L(\mathbb{X}) \: \underset{e\in\mathbb{E}}{\text{max}} \; \left\Vert e \right\Vert_2,
	 \end{split}
	 \end{aligned}
	 \end{equation}
	where $\mathbb{E}$ is a subset of $\mathbb{X}$ which includes the origin. This inequality defines a boxed shaped set that bounds the error in the nonlinear term. 
	\begin{equation}
	\mathbb{W}_g = \{ \zeta \in \mathbb{R}^n | \left\Vert \zeta \right\Vert_\infty \leq L(X) \underset{e\in\mathbb{E}}{\text{max}} \; \left\Vert e \right\Vert_2 \},
	\end{equation}		
	which can be added to $\mathbb{W}_m$ to make $ \mathbb{W} = \mathbb{W}_g \oplus \mathbb{W}_m$.	Basically, if a bound can be defined on the nonlinear term in the constraints region, then we can consider it as part of the additive disturbance.
	The next step is to find a bound for $\mathbb{W}_m$. To be able to use this method, all sources of uncertainty must be combined into a single additive disturbance. On major cause of uncertainty on this system is the delay in feedback loop due to $\tau_d$. This uncertainty can be bounded by finding the maximum state change that can happen in the maximum delay time. The following proposition explains the calculation of this bound.  
	\begin{proposition} 
		Let $x[k]\in X,u[k]\in U,a_{p}[k]\in A_p,w[k]\in W$ where all of the sets $X, U, A_p, W$ are bounded. Furthermore, assume that the radar and actuator delay is upper-bounded by $T_d$, i.e., $0\leqslant\uptau_d\leqslant T_d$. Then $w_\tau$, the uncertainty caused by delay, will be bounded by the set:
		\begin{equation*}\label{eq9}
		W_\uptau= T_d B_d K_c \times \left\{AX \oplus \left(BU\oplus (EA_p\oplus W)\right)\right\}.
		\end{equation*}
	\end{proposition}

	\begin{IEEEproof}: In order to prove this proposition, we use the fact that the difference between $x[k]$ and $x[k-\uptau_d]$ is given by the rate of changes of $x$ in $\uptau_d$-duration multiplied by $\uptau_d$ (assuming that $\uptau_d$ is small). Rigorously 
		
		\begin{equation*}\label{eq10}
		x[k-\uptau_d] =x[k]-\uptau_d \dot{x}[k].
		\end{equation*} 
		Next, note that according to \eqref{state_equation}, the set of all possible state change rates $\Delta X$ can be given by:
		\begin{equation*}\label{eq11}
		\begin{split}
		\Delta X= \{\dot{x} \mid \dot{x} =Ax+B u+B_p a_p+w, \\ \forall x \in X,\forall u\in U,\forall a_p\in A_p,\forall w\in W \},
		\end{split}
		\end{equation*}
		which, looking back at the preliminary definitions, is equivalent to the following Minkowski sum:
		\begin{equation*}\label{eq12}
		\Delta X=\left\{AX\oplus \left(BU\oplus \left(B_p A_p\oplus W\right)\right)\right\}.
		\end{equation*}
		Therefore, by equation \eqref{eq8}, the total amount of uncertainty that delay produces in the system is given by
		\begin{equation*}\label{eq13}
		W_\uptau= T_d B_d K_c \times \{AX\oplus \left(BU\oplus \left(B_p A_p\oplus W\right)\right)\},
		\end{equation*}
		which is what we aimed to show.
	\end{IEEEproof}
	Another source of uncertainty is the acceleration of the preceding vehicle. In (\ref{state_equation}) preceding vehicle's acceleration $a_p$ has been modeled as an additive disturbance. Therefore, $w_{a_p}$ which is the uncertainty caused by $a_p$ can be bounded by knowing a bound for maximum possible acceleration for the preceding vehicle. 
	\begin{equation}
	w_{a_p} \in \mathbb{W}_a = B_p A_p.
	\end{equation}
	Unknown model parameters can also increase model uncertainty. Uncertainty in vehicle mass, tire radius, drag coefficient, rolling resistance coefficients, road grade, wind speed and power-train efficiency must be considered in control design. If a bound for each of these uncertain parameters is available, equation (\ref{long_est_model}) can be used to find maximum model error that the uncertain parameters can cause. Suppose $ \Upsilon =\begin{bmatrix} m &r_w &C_d &\mu_{r_v} &\mu_{r_0} &\phi_r &v_w &\eta_p \end{bmatrix} $ as the vector of uncertain parameters in (\ref{long_est_model}) with $\overline{\Upsilon}$ as the vector their nominal value and $\Upsilon_{max}$ and $\Upsilon_{min}$ as the vector of their maximum and minimum values. Then the following optimization problem will find the maximum model error. 
	\begin{equation*}
	\begin{aligned}
	\begin{split}
	 e_{a_{min}} &= \underset{\Upsilon, v_h, T_{com}} {\min}  {a_h(\Upsilon,v_h,T_{com})} - {{a}_h(\overline{\Upsilon},v_h,T_{com})}, \\
	 e_{a_{max}} &= \underset{\Upsilon, v_h, T_{com}} {\max}  {a_h(\Upsilon,v_h,T_{com})} - {{a}_h(\overline{\Upsilon},v_h,T_{com})},\\
	  \text{subject to}\\
	 &\qquad \Upsilon_{min} \leq\Upsilon \leq \Upsilon_{max},\\
	 &\qquad  0 \leq v_h \leq v_{h_{max}},\\
	 &\qquad  T_{min} \leq T_{com} \leq T_{max}.
	 \end{split}
	\end{aligned}
	\end{equation*}
	where $e_{a_{min}}$ and $e_{a_{max}}$ are the minimum and maximum error caused by parameter uncertainty which based on them the set of all possible acceleration errors due to parameter uncertainty can be defined as: $e_a \in E_a$ and bounded additive disturbance due to the parameter uncertainty can be calculated
	\begin{equation}
	w_{a_h} \in \mathbb{W}_h = Bg E_a
	\end{equation}
	where $B_g$ is as defined in (\ref{state_equation}). Combination of all the uncertainty sources will be the bounded additive disturbance term. 
	\begin{equation}
	\mathbb{W}=\mathbb{W}_g \oplus \mathbb{W}_\uptau \oplus \mathbb{W}_a \oplus \mathbb{W}_h 
	\end{equation}
	This disturbance set will be used for the design of the tube-based controller. (\ref{error_model}) can be rewritten as
	\begin{equation}
	e_[k+1]= A_c e[k] + w[k]  \qquad  w\in\mathbb{W}.
	\end{equation}
	Using this stable model with a bounded additive disturbance and Minkovski sum, the finite reachable set for the error can be calculated. 
	\begin{equation}
	\Phi_{n} =  \oplus^{n}_{i=0}{A_c}^i \mathbb{W}
	\end{equation}
	where $\Phi_n$ is the finite reachable error set and it's infinity limit $\Phi_{\infty}$ is called the robust positive invariant set \cite{kolmanovsky1998theory}. In this paper, our T-MPC is similar to \cite{chisci} which use finite invariant set instead of infinity RPI set with fixed current state. 
	\subsection{Model adaptation}	
	A model adaptation method is employed to adapt to changes in the system and environment in order to maintain the performance of designed controllers. In this paper, we use a least square parameter adaption method with forgetting factor similar to \cite{ioannou2006adaptive} with same notation. This method uses previously presented parametric models, to estimate the value of each effective parameter. It works based on minimizing the squared error between the estimated and measured output of the system by minimizing the following cost function. 
	\begin{equation}
	\begin{aligned}
	\begin{split}
		J(\hat{\theta})=  \frac{1}{2}\int_{0}^{t} &\frac{e^{-\beta(t-\tau)} (z(\tau)-\hat{\theta}^T (t)\phi(\tau))^2}{m^2_s (\tau)} d\tau \\ &+ \frac{1}{2} e^{-\beta t} (\hat{\theta}(t) -\hat{\theta}_0)^TQ_0(\hat{\theta}(t)-\hat{\theta}_0),
	\end{split}
	\end{aligned}
	\end{equation}
	where $\hat{\theta}$ is the estimated parameters vector, $\hat{\theta}_0$ is the initial estimated parameter, $\phi$ is the measured input signal, $z$ is the measured output signal, $\beta$ is a forgetting factor, $Q_0$ is a weighting matrix, $P$ is covariance matrix and $m^2_s$ is a normalizing term that can be chosen as: $m^2_s = 1+\alpha \phi^T\phi, \quad  \alpha \geq 0$. By minimizing this cost function, the algorithm can find an estimation of the parameters. The first term penalizes the estimation error and the second term penalizes the convergence rate with a decaying factor in order to increase estimation robustness against disturbances. Forgetting factor gives a higher weight to the new measurements, so that in case of change in a parameter the algorithm can adapt to it. Based on this cost function a recursive least square algorithm is defined as follows:
	\begin{equation}
	\begin{aligned}
	\begin{split}
	&\dot{\hat{\theta}}(t) = P(t)\epsilon(t)\phi(t),\\
	&\dot{P}(t)=\beta P(t) - P(t) \frac{\phi(t)\phi^T(t)}{m^2_s(t)}P(t),\\
	&\epsilon(t)= \frac{z(t)-\hat{\theta}^T(t)\phi(t)}{m^2_s(t)}.
	\end{split}
	\end{aligned}
	\end{equation}
	This algorithm updates the covariance and estimated parameters on-line when the vehicle is running. To prevent wrong estimation, it is necessary to limit the estimated parameters. Therefore, parameter projection was used to put constraints on estimations. Moreover, to avoid the covariance matrix from becoming very large, it is also necessary to put a constraint on its maximum value. Assuming the desired constraint on the parameters is defined by: $ S=\{\theta\in R^n | g(\theta)\leq 0 \} $, where $g$ is a smooth function and $ R_0 $ as an upper bound for $ P $, projection can be defined as follows:
	\begin{equation}
	\begin{split}
	\begin{aligned}
	&\dot{\theta}=\begin{cases}
	P\epsilon \phi & \text{if} \: \theta \in S^o \: \text{or}\\
	& \: \theta \in \delta(S) \: \& \: (P\epsilon\phi)^T \nabla g \leq 0 \\
	P \epsilon \phi - P \frac{\nabla g \nabla g^T}{\nabla g^T P \nabla g} P \epsilon \phi & otherwise
	\end{cases} \\
	&\dot{P}=\begin{cases}
	\beta P - P \frac{\phi\phi^T}{m^2_s}P & \text{if} \: \Vert P \Vert \leq R_0 \; \& \{ \theta \in S^o \; \text{or} \\
	& \: \theta \in \delta(S) \: \& \: (P\epsilon\phi)^T \nabla g \leq 0 \} \\
	0 &otherwise
	\end{cases}
	\end{aligned}
	\end{split}
	\end{equation}
	Projection ensures that the estimation will not go out of the constraint region and will move along the border when it reaches to its limits. This adaptation algorithm is used estimate fuel consumption, electricity consumption and longitudinal dynamics parameters based on the reduced model presented in the modeling section. The estimated parameters will be used in the control-oriented model of AT-NMPC so that the optimization problem will find the updated optimal point of the system.
	
	\subsection{Adaptive robust controller}
	Using the disturbance set and the parameter adaption method above, it is possible now to define our adaptive robust control problem. This controller includes a linear controller that stabilizes the system, and an NMPC that controls the system based on its nominal control-oriented model without considering the uncertainty and disturbances. NMPC keeps the nominal state of the system in a tighter region to ensure that the actual system states will remain inside the defined constraints. Moreover, parameter adaptation updates the control-oriented model that is used in the definition of the cost function to maintain system performance. Therefore, two control-oriented models are used here, one for updating the cost function and performing a future prediction and another one for handling of the constraints.
	
	\begin{equation}\label{ATBMPC_problem}
	\begin{aligned}
	\begin{split}
	\underset{c_{o}} {\min} \Big\{ \sum_{i=1}^{Np} & \big( \omega_1 e_p^2(\hat{x},\hat{u}) + \omega_2  e_v^2(\hat{x},\hat{u}) + \omega_3 \hat{u}^2 \\ & \quad \quad \quad  +\omega_4 E_{cost}(\hat{x},\hat{u},PR,\hat{A}, \hat{\Gamma}) \big) \Big\},\\
	\text{subject to} \\
	& \bar{x}[n]=x[n], \quad \hat{x}[n]=x[n],\\
	&\hat{x} [n+i+1]= \hat{f}_n(\hat{x}[n+i],c_0[n+i]),\\
	&\bar{x} [n+i+1]= \bar{f}_n(\bar{x}[n+i],c_0[n+i]),\\
	& \hat{u}[n+i] = -K_c\hat{x}[n+i]+c_0 [n+i], \\
	&\bar{x} [n+i+1]\in X\ominus {\Phi}[i], \\
	&c_{0} [n+i+1]\in U\ominus(-K_c {\Phi}[i]),
	\end{split} 
	\end{aligned}
	\end{equation}
	where $\bar{f}$ and $\hat{f}$ are the nominal and estimated nonlinear longitudinal dynamic model of the vehicle, $N_p$ is the prediction horizon's length, $\omega_1$, $\omega_2$, $\omega_3$ and $\omega_4$ are weights on each term and other parameters are as defined before. The control problem finds a vector $c_0$ that minimizes the cost function in the prediction horizon while the actual input to the system is combination of $c_0$ and the linear controller. 
	
	\remark In this control problem, the current state of the system is not a decision variable and $x[n]$ has a fixed value. Both nominal and adapted control-oriented models start from the same initial point but perform future predictions based on their own parameters. 
	\remark The tighter constraints on the states ensures that the system states would remain inside $X$, the defined state constraint, for any amount of disturbance that satisfies $w \in \mathbb{W} $. Moreover, the tighter constraints on the input reserves a part of available actuation for the linear controller to maintain system's robustness. 
	\remark  Having two separate control-oriented models ensures that constraints are always satisfied based on the fixed nominal control-oriented model. Therefore even if the adapted control-oriented model has low accuracy, the system's robustness will be maintained. This is specially important in the case that a sudden change in model parameters occur. Because the parameter estimator may not be able to recognize the change in the model immediately, it is necessary to make sure that the system will remain safe and robust while the model is getting adjusted which is achievable by using separate models as has been done here.
	
	\subsection{Fast optimizer}
	To implement the proposed robust Eco-ACC on a vehicle control system, the AT-NMPC problem must be solved in real-time. Therefore a fast solver is required that can solve the nonlinear optimization problem with little computational demand. To this end, in this paper, Newton/GMRES method has been used to solve the control optimization problem. This method is claimed to be very fast as it solves the differential equation once at each time step \cite{kelley1999iterative}. In the current study, the authors use an automatic multi-solver NMPC code generator, called MPSee, to generate the NMPC code based on the Newton/GMRES algorithm which has been previously developed and tested in the authors' research group \cite{tajeddin2016automatic},\cite{tajeddin2017ecological}. MPSee is a MATLAB-based mathematical program that enables users to develop GMRES-based NMPC codes for different optimal control problems and carry out simulations in Simulink. To define the Newton-GMRES solver, field vector and constraints have been defined as follows:
	\begin{equation}\label{GMRES}
	\begin{aligned}
	\begin{split}
	&f(x,u)=\frac{d}{dt}(\begin{bmatrix}
	\bar{p}_h & \bar{v}_h & \bar{T}_w & \bar{p}_p & \bar{v}_p & \bar{a}_p & \hat{p}_h & \hat{v}_h & \hat{T}_w
	\end{bmatrix}^T) \\ &= \begin{bmatrix}	
	\bar{v}_h \\
	 \frac{\bar{T}_w}{m r_w \eta_p}-\{ \frac{\frac{1}{2} \rho_a C_a A_w}{m} \bar{v}_h^2 + g \bar{\phi}_r + \mu_v g \bar{v}_h +g \mu_0 \} \\	
	 -\frac{\bar{T}_w}{\tau_a}+\frac{u-K[\bar{e}_p \;\; \bar{e}_v \;\; \bar{T}_w]}{\tau_a} \\
	 \bar{v}_p \\
	 \bar{a}_p \\
	 - \sigma \bar{a}_p \\
	 \hat{v}_h \\
	 \Theta_1 \hat{T}_w -\Theta_2 \hat{v}_h^2 -\Theta_3 \hat{\phi}_r - \Theta_4 \hat{v}_h -\Theta_5\\
	 -\frac{\hat{T}_h}{\tau_a}+\frac{u-K [ \hat{e}_p \;\; \hat{e}_v \;\; \hat{T}_h]}{\tau_a}
	\end{bmatrix}
	\end{split}
	\end{aligned}
	\end{equation}
	\begin{equation}\label{con_GMRES}
	\begin{aligned}
	\begin{split}
	C(x,u)=	H_x[i] \begin{bmatrix}
	\bar{e}_p[i] \\ \bar{e}_v[i] \\ u[i]-K [\bar{e}_p [i] \;\; \bar{e}_v[i] \;\; \bar{T}_w[i] ] 
	\end{bmatrix}-h_x[i]
	\end{split}
	\end{aligned}
	\end{equation}
	where $H_x[i]$ and $h_x[i]$ are used to define the polytopic constraints in each step equivalent to $X\ominus \Phi[i]$ in (\ref{ATBMPC_problem}). The two separate control-oriented models have been implemented in here in the field vector and constraints where barred variables show the nominal values and hatted variables are the estimated ones. $\sigma$ is the decaying factor for preceding vehicle's acceleration as defined in \cite{sakhdari2016ecological} .
		\begin{figure}
		\centering
		\makeatletter
		\subfloat{\includegraphics[width = 3.2in]{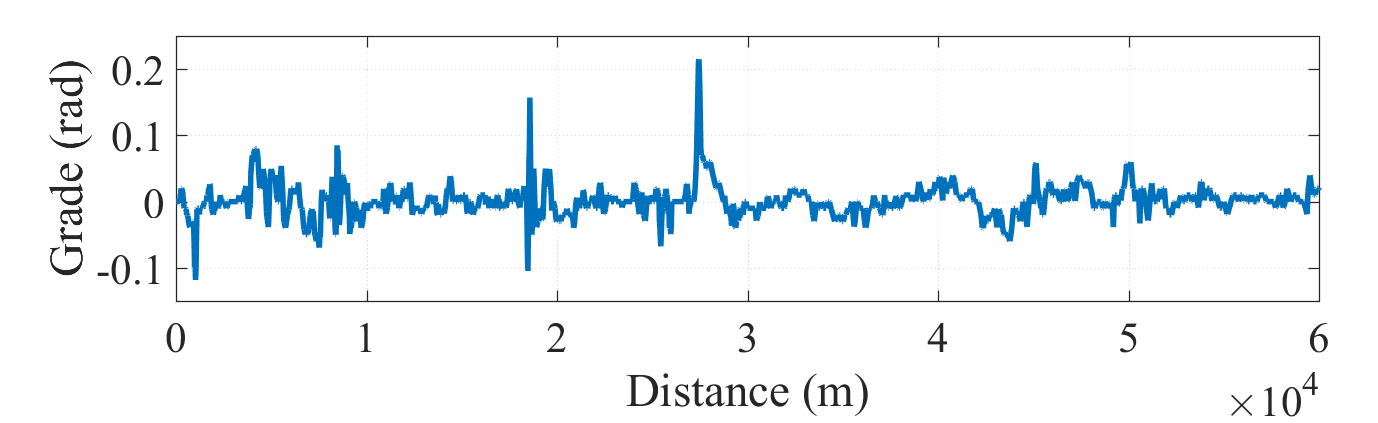}} \\
		\vspace{-0.3cm}
		\subfloat{\includegraphics[width = 3.2in]{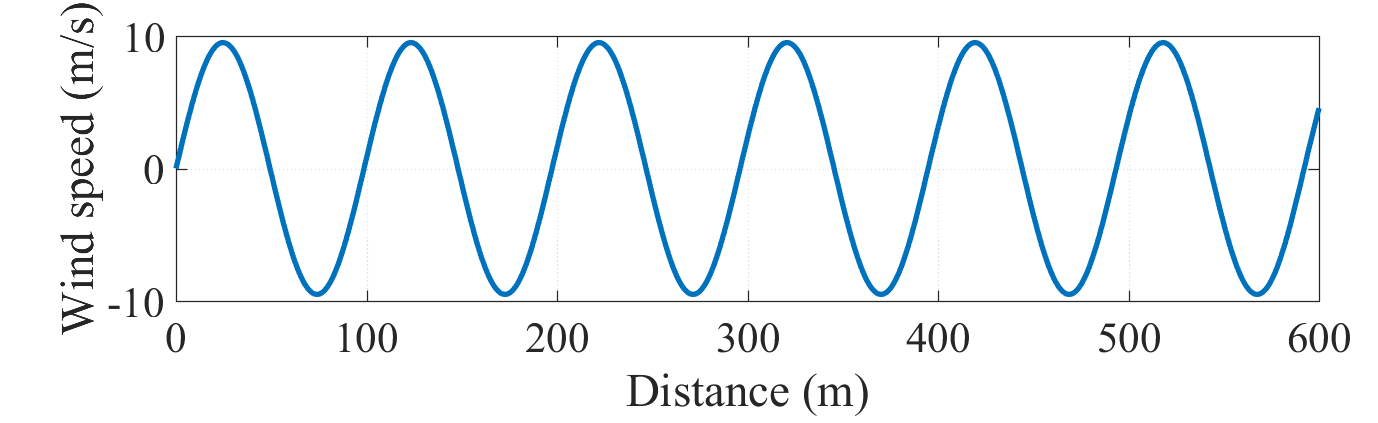}} \\
		\caption{grade and wind profile injected in the simulation environment}
		\vspace{0.35cm}
		\label{fig:windgrade}
	\end{figure}
	\begin{figure}
		\centering
		\makeatletter
		\vspace{-0.3cm}
		\subfloat{\includegraphics[width = 1.75in]{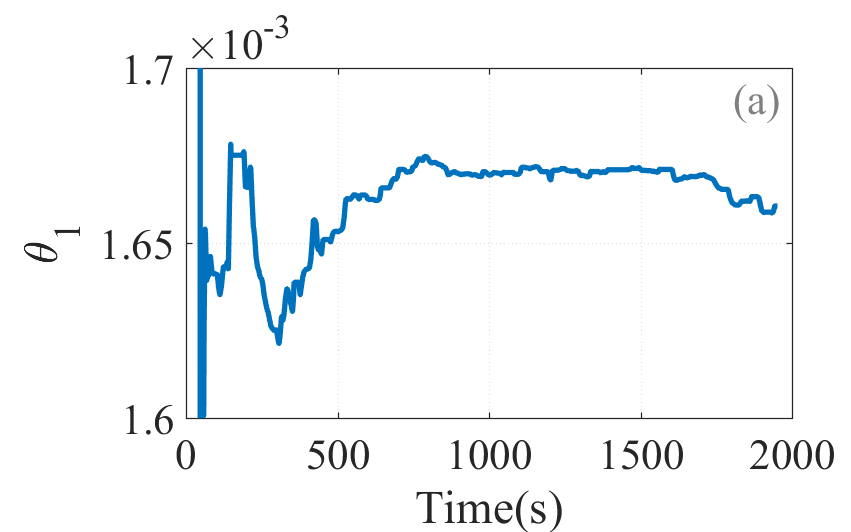}} 
		\subfloat{\includegraphics[width = 1.75in]{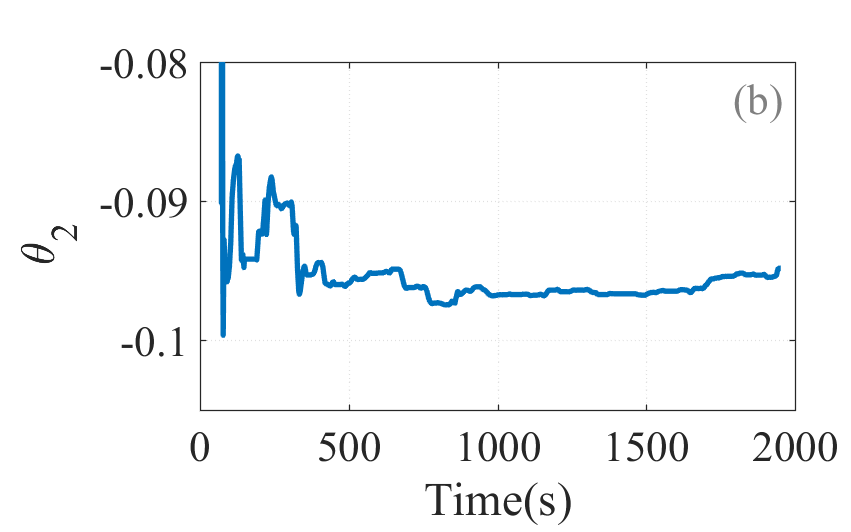}} \\
		\vspace{-0.3cm}
		\subfloat{\includegraphics[width = 1.75in]{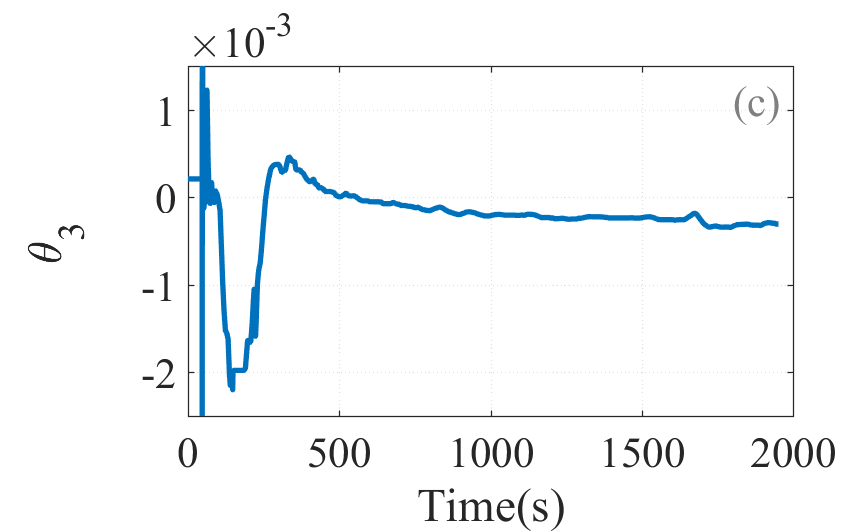}} 
		\subfloat{\includegraphics[width = 1.75in]{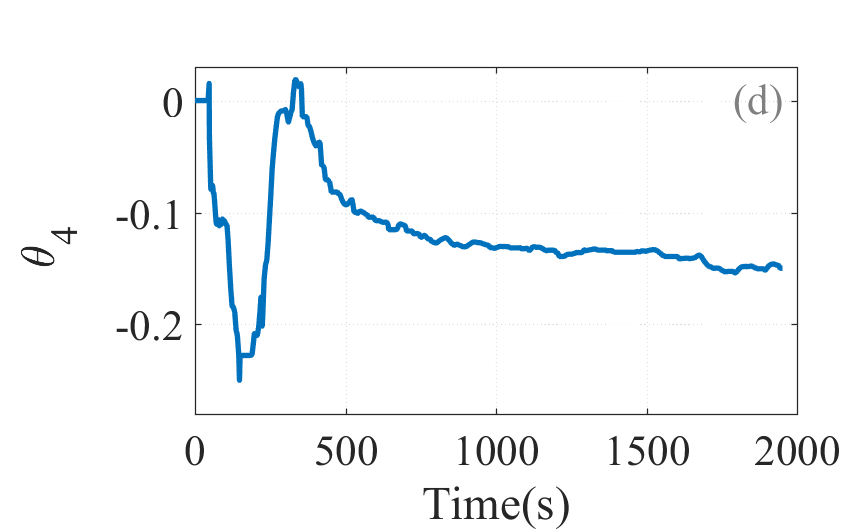}} \\
		\vspace{-0.3cm}
		\subfloat{\includegraphics[width = 1.75in]{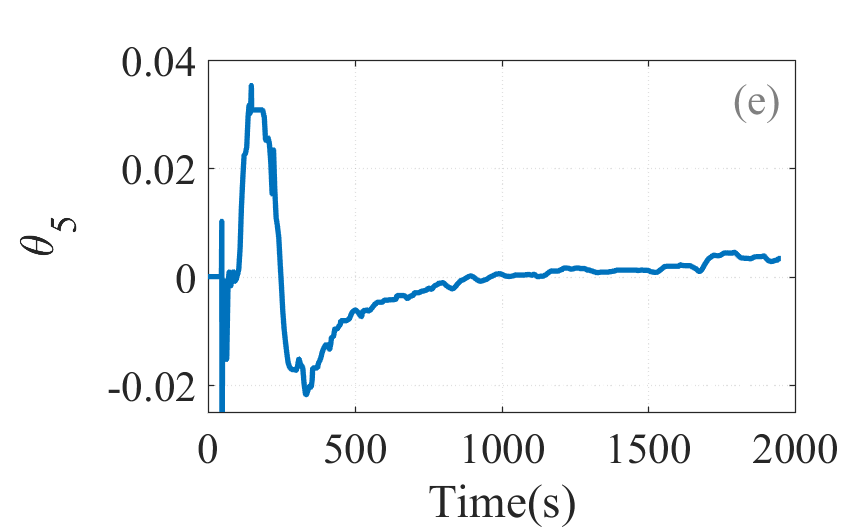}}
		\subfloat{\includegraphics[width = 1.75in]{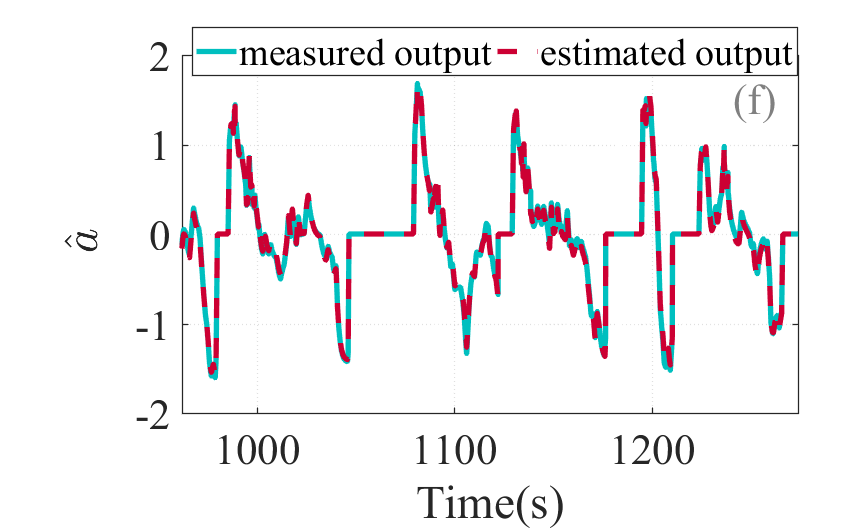}} \\
		\caption{Longitudinal dynamic parameter estimator}
		\vspace{0.25cm}
		\label{fig:theta_long}
	\end{figure}
	\section{Control evaluation}
	This section presents the evaluation of the proposed Eco-ACC in terms of estimation, robustness and ecological improvement. A high-fidelity model of the base-line vehicle, developed in Autonomie, is used for evaluation tests. First, the performance of the three estimators is presented. Second, robustness of the proposed controller is tested using high-fidelity model of the base-line vehicle. Third, ecological improvement caused by the the proposed Eco-ACC is discussed. Finally, the result of HIL experimet is presented that shows the real-time capability of the proposed controller.
	\begin{figure}
	\centering
	\makeatletter
	\vspace{-0.3cm}
	\subfloat{\includegraphics[width = 1.75in]{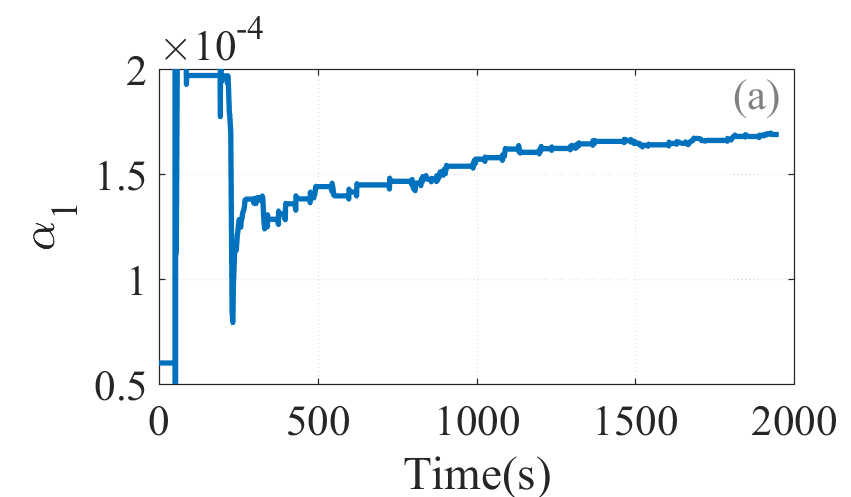}} 
	\subfloat{\includegraphics[width = 1.75in]{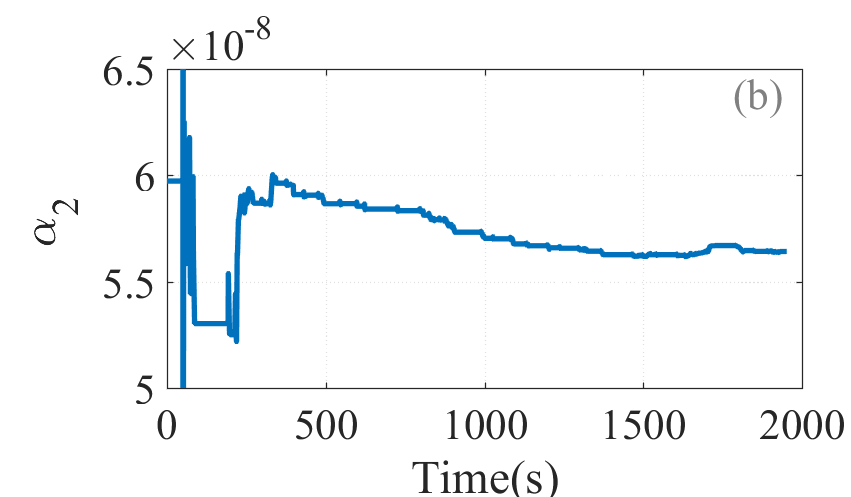}} \\
	\vspace{-0.3cm}
	\subfloat{\includegraphics[width = 1.75in]{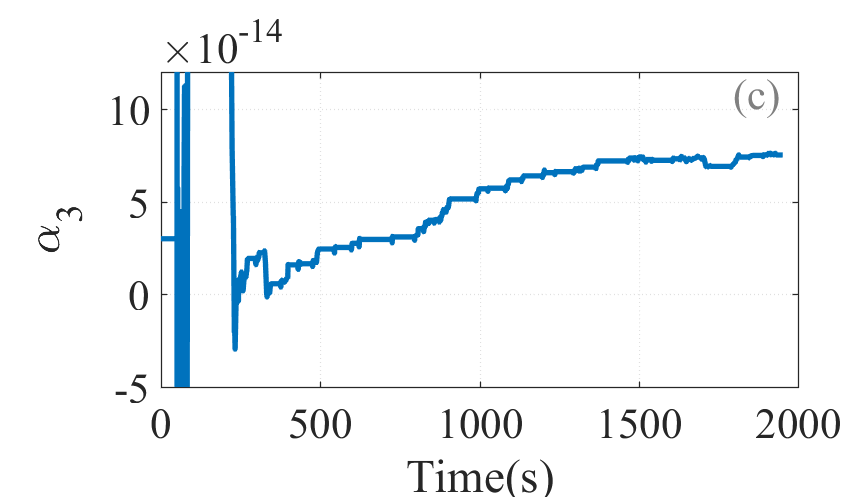}} 
	\subfloat{\includegraphics[width = 1.75in]{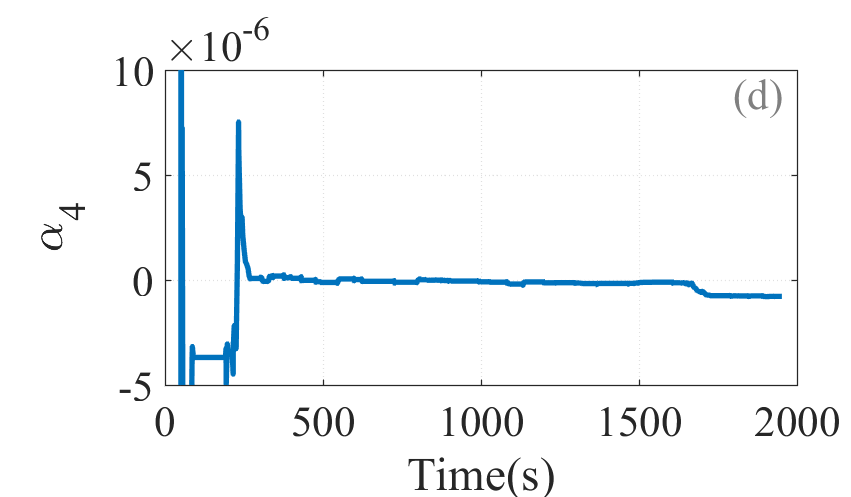}} \\
	\vspace{-0.3cm}
	\subfloat{\includegraphics[width = 3.25in]{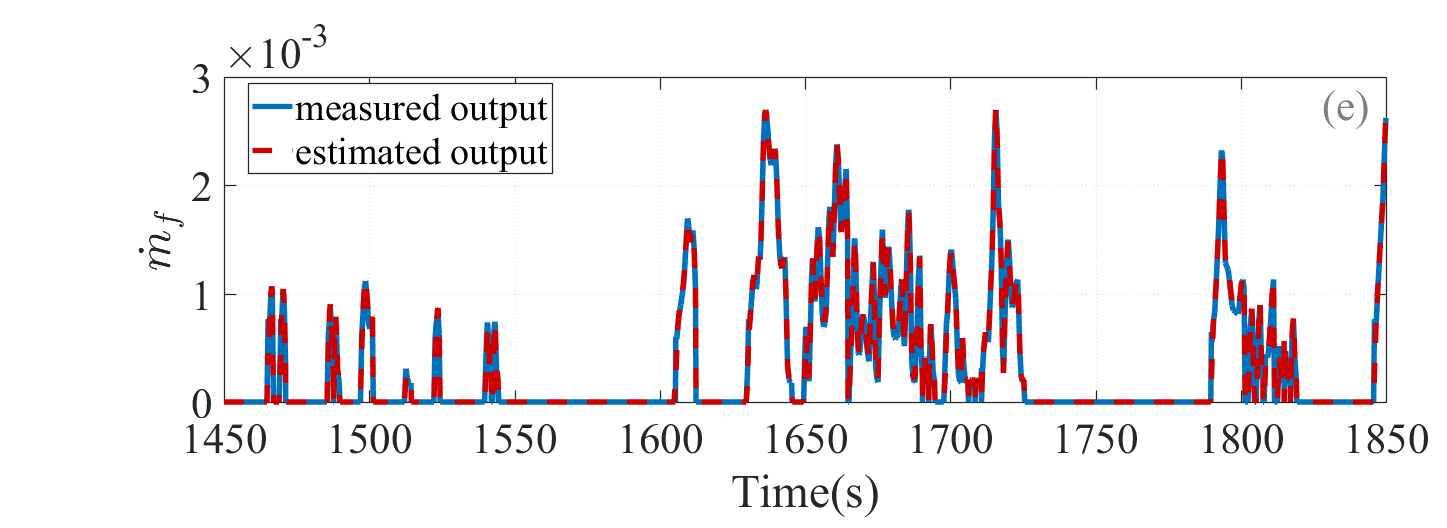}}
	\caption{Fuel consumption parameter estimator}
	\label{fig:theta_eng}
\end{figure}
\begin{figure}
	\centering
	\makeatletter
	\vspace{-0.3cm}
	\subfloat{\includegraphics[width = 1.75in]{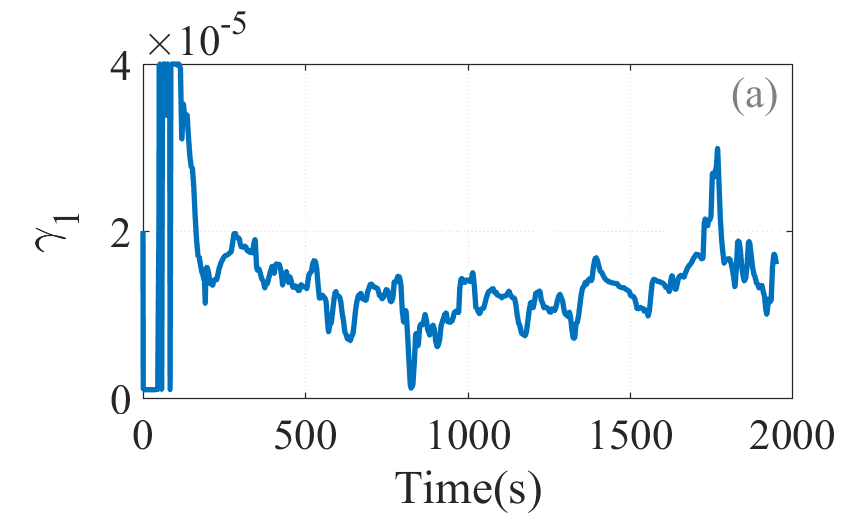}} 
	\subfloat{\includegraphics[width = 1.75in]{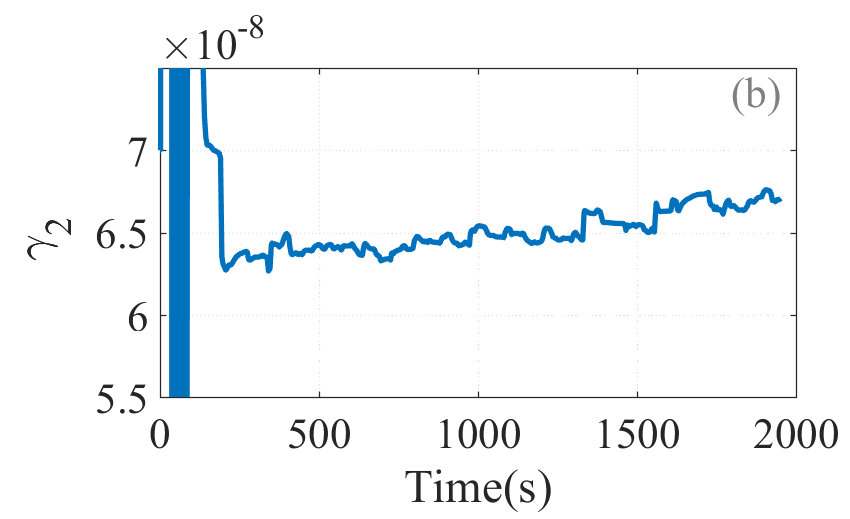}} \\
	\vspace{-0.3cm}
	\subfloat{\includegraphics[width = 1.75in]{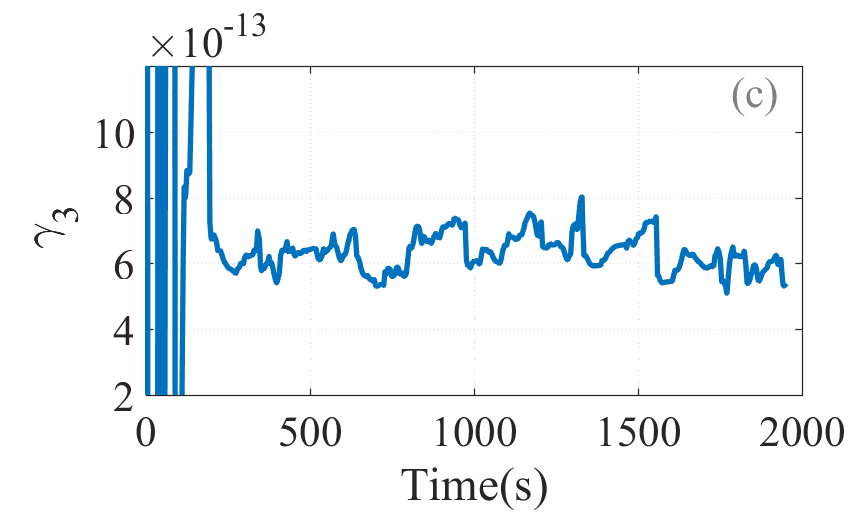}} 
	\subfloat{\includegraphics[width = 1.75in]{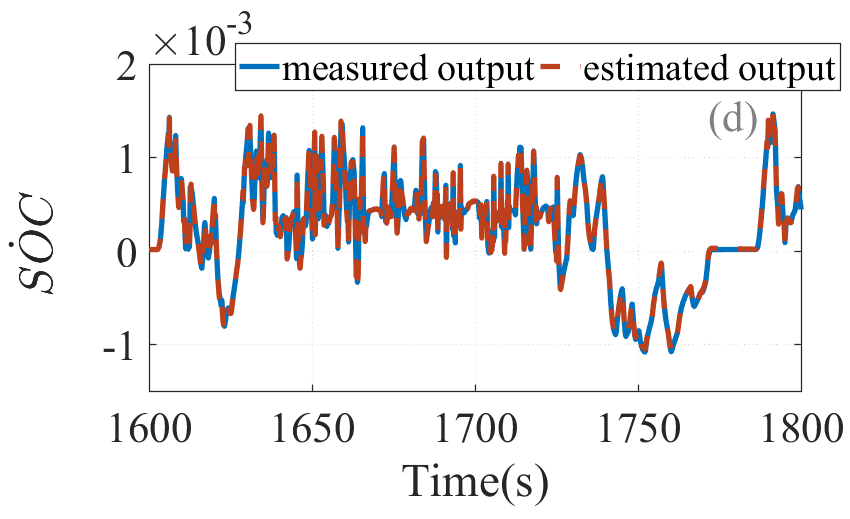}} 
	\caption{Electricity rate parameter estimator}
	\label{fig:theta_mot}
\end{figure}
	\subsection{Parameter estimation}
	We used three least-square parameter estimators to improve the control-oriented model of our  predictive controller. The first estimator gets velocity, road grade and propulsion or braking torque, and then finds the parameters of the longitudinal model based on (\ref{parametrimodel_long}). Fig. \ref{fig:theta_long} (a-e) illustrate the result of on-line parameter estimation for this estimator and Fig. \ref{fig:theta_long} (f) shows the estimated acceleration by the estimated model compared to the measured acceleration. It is worth noting that the estimated parameters may not converge to their real value which, for adaptive control use, is acceptable as long as the estimated output matches the real value \cite{ioannou2006adaptive}. This model predicts the motion of the vehicle inside the prediction horizon and adjusts to changes, so that the prediction would be more accurate. The two other estimators are shown in Fig. \ref{fig:theta_eng} for fuel consumption and in Fig. \ref{fig:theta_mot} for electricity. As it can be seen both models are able to estimate the desired output closely and adapt to changes in the parameters. The initial guess for the estimators was chosen arbitrary, which is the reason for initial oscillations in the parameter estimations. To improve the estimators' performance, forgetting factor has higher value in the beginning and decreases gradually afterward to a lower value. The main objective of using estimators is for making sure that the optimizing the defined cost function would optimized the actual system. Therefore, the result of these estimators will be used in the cost function of the optimization problem.

	\subsection{Robust constraint handling}
\begin{figure}
		\centering
		\makeatletter
		\vspace{-0.3cm}
		\subfloat{\includegraphics[width = 3.25in]{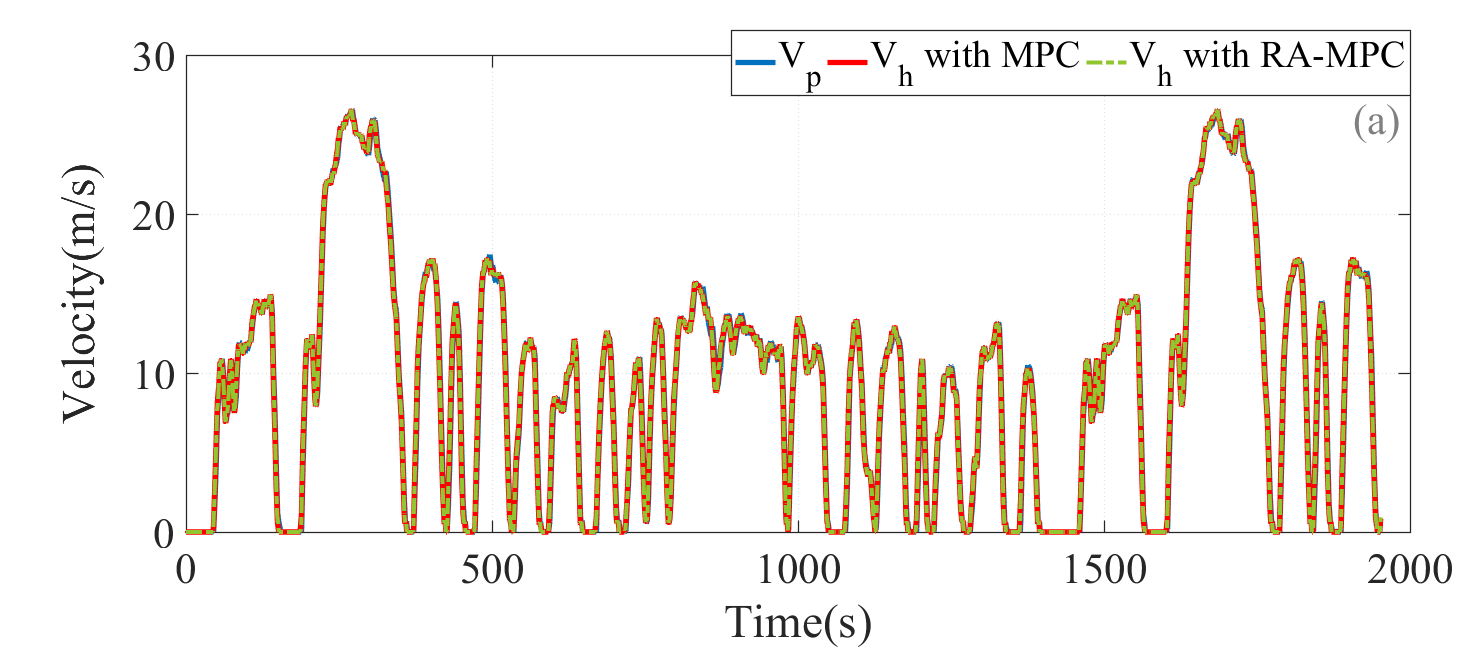} } \\
		\vspace{-0.3cm}
		\subfloat{\includegraphics[width = 3.25in]{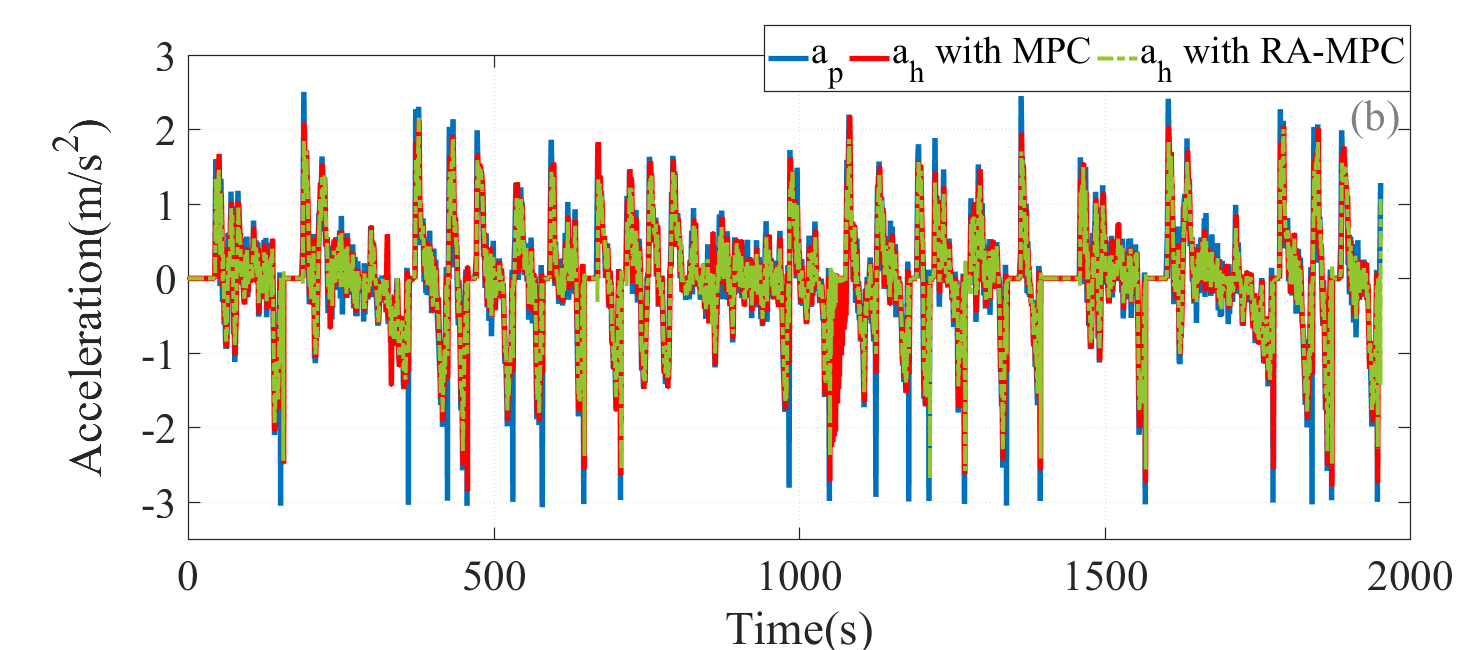} } \\
		\vspace{-0.3cm}
		\subfloat{\includegraphics[width = 3.25in]{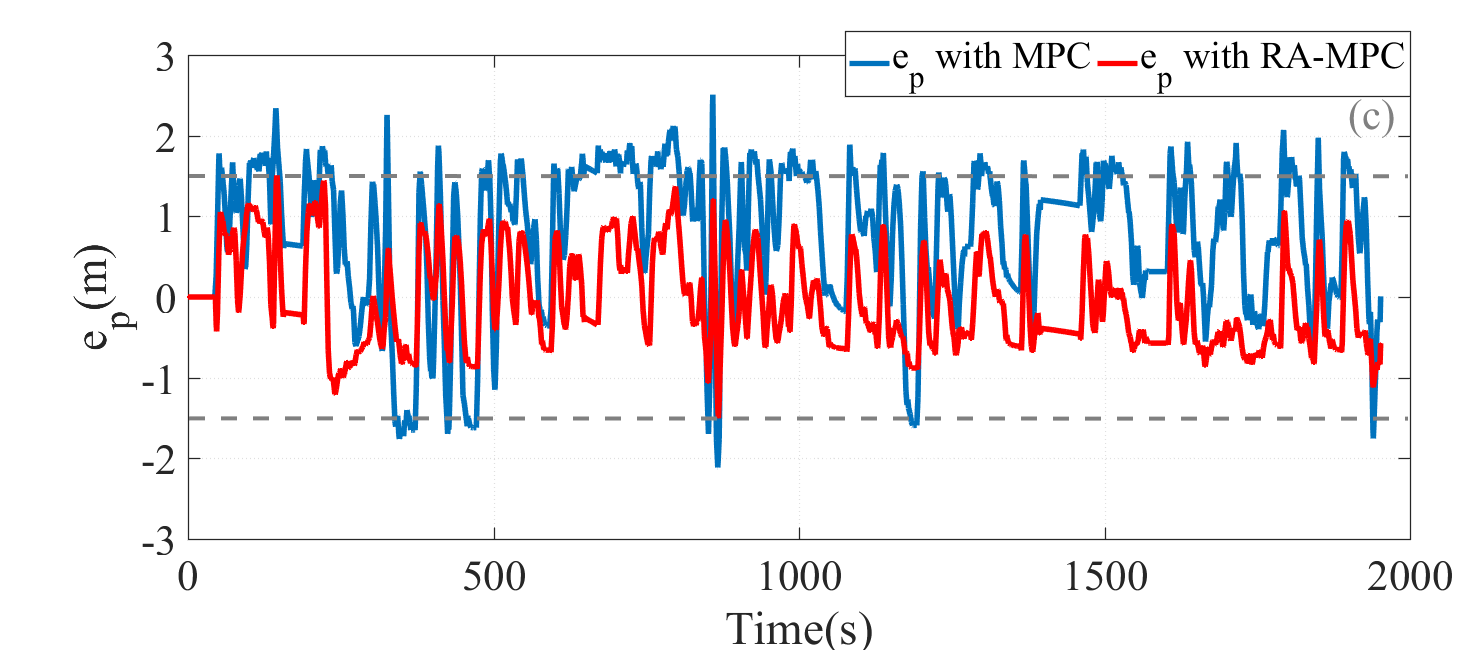} } 
		\caption{(a) velocity, (b) acceleration and (c) position error in car-following simulation}
		\label{fig:const_hndl}
	\end{figure}

	Other than on-line parameter adaptation, the proposed AT-NMPC based Eco-ACC is able to handle bounded uncertainties. To show the validity of this statement, we conducted simulations by utilizing a high-fidelity model of the base-line vehicle and then added a variable wind speed and road grade to the simulation environment, as shown in Fig. \ref{fig:windgrade}. During the simulation, host vehicle follows a preceding vehicle in a drive cycle by receiving inter-vehicular distance and velocity from the radar. To simulate the effect of delay in the radar's data, 400ms transport delay was injected in the inter-vehicular distance and velocity during the simulation. Moreover, parametric errors were considered in the control-oriented model by 20\% error in the vehicle mass, 50\% error in the drag coefficient, and ignoring the rolling resistance forces. Then based on the method presented in section IV, disturbance sets were calculated and used for defining the constraints of the AT-NMPC problem. Fig. 8 shows the result of the simulation in a standard FTP-75 drive cycle. The preceding vehicle follows the given drive cycle and the host vehicle follows the preceding vehicle during the simulation using a regular MPC based ACC and proposed AT-NMPC Eco-ACC. As shown in Fig. \ref{fig:const_hndl} (a) both controllers have acceptable velocity tracking performance while following the preceding vehicle. However, the NMPC based ACC has harsher accelerations compared to the proposed Eco-ACC. Harsher accelerations are due the fact that non-robust NMPC can not handle the defined constraints due to the existed uncertainties in the system. Therefore, these uncertainties can push the system out of the defined constraints and NMPC has to perform harsh braking and acceleration to go back into the constraints. However, AT-NMPC is robust against these uncertainties and therefore it has less harsh accelerations.  Fig. \ref{fig:const_hndl} (c) compares the position tracking error of the two controllers. NMPC is not able to handle the defined constraints and it has higher position error than the defined limits. On the other hand, AT-NMPC handles the constraint perfectly because the effect of uncertainties has been considered in its design based on the optimization problem in (\ref{ATBMPC_problem}).

	\subsection{Ecological Improvement}
\begin{figure}
		\centering
		\makeatletter
		\vspace{-0.3cm}
		\includegraphics[width = 3.25in]{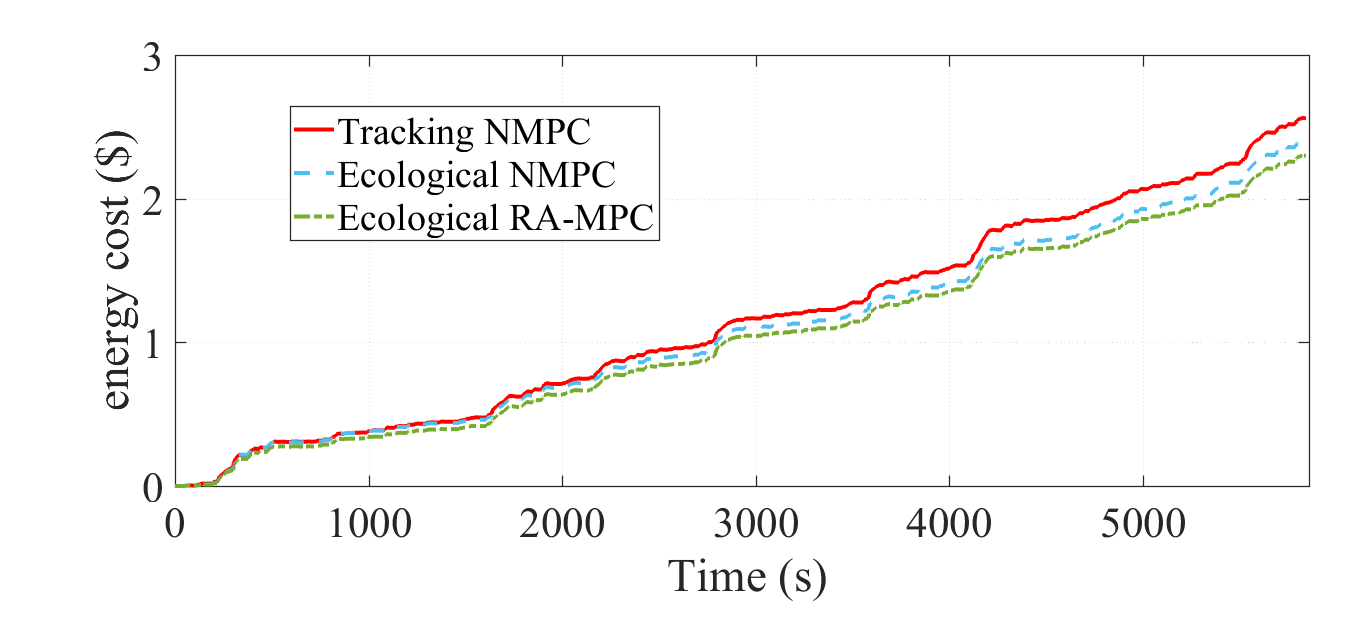}  \\
		\caption{Energy consumption in three consecutive FTP-75 drive cycles}
		\label{fig:ecological}
	\end{figure}

	The objective function of the proposed controller in defined to minimize the cost of energy in a drive cycle. Based on the given dynamic model of the vehicle and available road elevation, AT-NMPC predicts the future host vehicle’s trajectory and based on that calculates the expected power demand in the prediction horizon. Then, based on the power ratio of energy management system, it calculates the power demand of each energy source and also energy cost during the prediction horizon. Parameter estimators make sure that the optimal point of the cost function is the actual optimal point of the system. Fig. \ref{fig:ecological} compares the trip energy cost of a tracking NMPC, ecological NMPC and the proposed AT-NMPC in three consecutive FTP-75 drive cycles. A longer drive cycle is chosen to minimize the effect of energy management in the achieved results. Tracking NMPC has higher weightings on position and velocity tracking term to increase the tracking performance. Therefore, it mostly sacrifices energy cost for better tracking and has the highest energy cost in this driver cycle by \$2.56. The Eco-NMPC has a higher weighting on energy cost term, which means that it sacrifices tracking, inside the defined constraints, to have better energy cost. As such, it has an energy cost of \$2.4, which is about 6.2\% lower than tracking NMPC. However, in this case, NMPC controller is not able to handle the defined constraints and uncertainties push the system out of the defined constraints set. To be able to move back into the constraints, NMPC has to do harsh braking and accelerations, which increase the energy cost. AT-NMPC, on the other hand, handles the defined constraints and also minimizes a cost function that adapts to the actual vehicle behavior. It has an energy cost of \$2.29, which is about 10.5\% lower compared to tracking NMPC. 

	\subsection{Real-time Implementation}
	To further examine the performance of the proposed controller, hardware-in-the-loop (HIL) experiments have been conducted to study the potential and capability of AT-NMPC for real-time implementation in a vehicle control system. HIL tests take into account the computational limits and communication issues, and their result is considered more practical than software-in-the-loop simulations. Because physical prototyping in the early stages of development of vehicle control systems could be very expensive, HIL experiments, which are less expensive and also faster and safer, usually carried out before manufacturing the prototype vehicle \cite{Golchoubian}. In this paper, dSPACE Micro-Autobox II control prototyping hardware was used for HIL tests. This setup is one of the widely used instruments for calibration and testing of ECUs specifically for automotive applications. As shown in Fig.10, the HIL setup has three main components: 1) a prototype ECU (MicroAutoBox II), which is an independent processing module that runs the uploaded controller; 2) a real-time simulator (DS1006 processor board), which is responsible for running the complex high-fidelity model of the vehicle in real-time fashion; and 3) a personal computer (PC) that serves as the human-machine interface and is used for programming the real-time machine and prototype ECU, as well as for recording the desired test signals. All the communications between the prototype ECU and the real-time simulator are performed through a Controller Area Network (CAN).
	
	For HIL tests, a C code was generated of the designed controller by using the dSPACE Real-Time Workshop code generator and uploaded to the prototype ECU. With a similar procedure, the high-fidelity model was uploaded to the realtime simulator using the human-machine interface. Fig.11 illustrates the turnaround time of the controller in FTP-75 drive cycle simulation. The turnaround time for this controller is between 400$\mu s$ and 700$\mu s$ at all times for a prediction horizon of $N_p = 10$. The maximum inner and outer iterations set to be less than 5 to enable real-time implementation.
	\begin{figure}
		\centering
		\makeatletter
		\vspace{-0.3cm}
		\includegraphics[width = 3.25in]{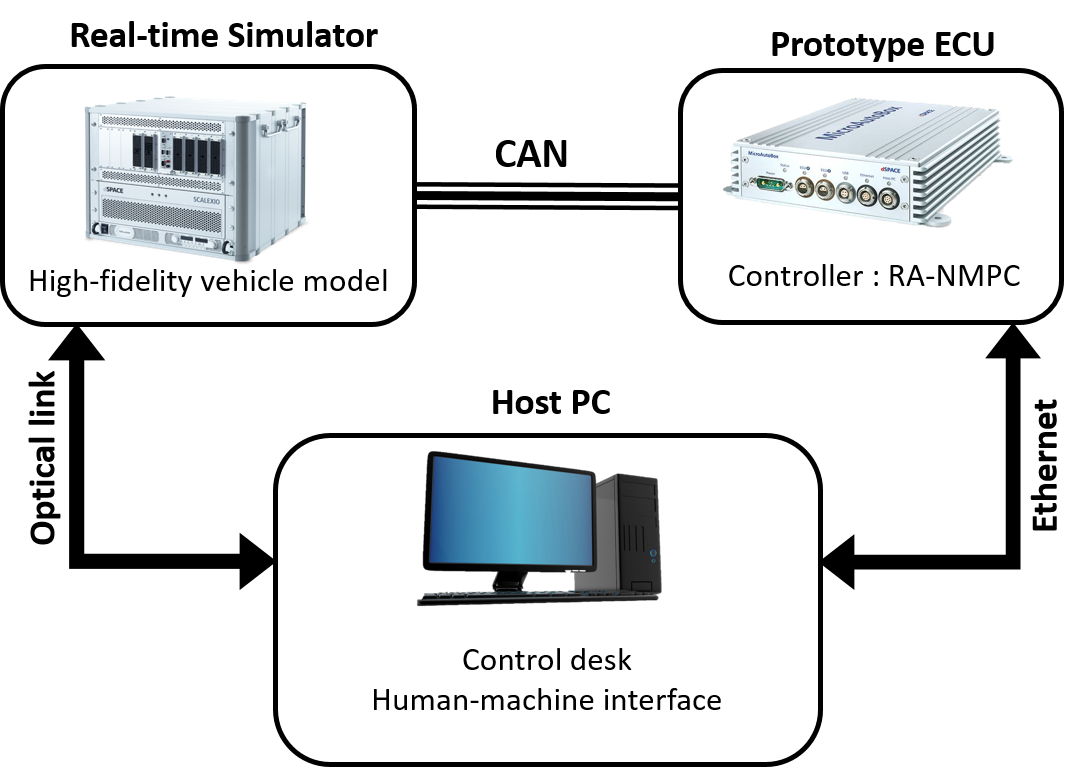}  \\
		\caption{Schematic of the HIL experiment setup}
		\label{fig:HILsetup}
	\end{figure}
	\begin{figure}
		\centering
		\makeatletter
		\vspace{-0.3cm}
		\includegraphics[width = 3.25in]{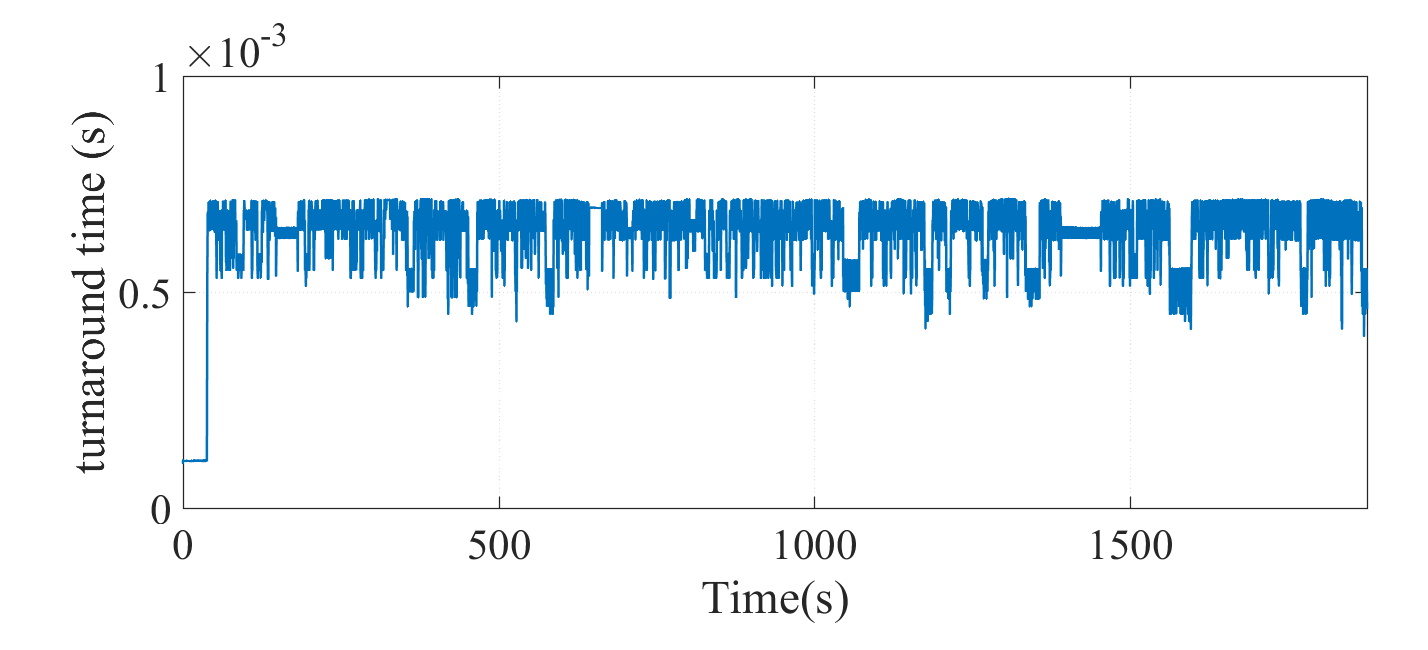}  \\
		\caption{Turn-around time of the controller in HIL experiment}
		\label{fig:HIL}
	\end{figure}

	\section{Conclusions}
	This paper proposed an adaptive tube-based nonlinear model predictive controller for the design of autonomous cruise control systems. This method ensures the robust satisfaction of the defined constraints in the presence of uncertainty, and also improves the system’s performance by adapting to the changes in the vehicle control-oriented model. Therefore, in a way, this method decouples performance and robustness by using separate models one for constraint handling and another one for defining the objective function.

	In the modeling step, a nonlinear control oriented model was presented for a vehicle that performs car-following. This model was used for evaluation of the safe sets in the presence of additive disturbance. Moreover, models for fuel consumption and electricity rate were presented to estimate the cost of energy in the prediction horizon and based on them, reduced models for parameter estimation were generated. A high-fidelity model of the base-line PHEV, Toyota plug-in Prius, was used to evaluate the controller.

	In the control design step, first, a linear controller was used to stabilize the system. Then by analyzing the existed uncertainties, they were translated into an additive disturbance term to define a bound for maximum uncertainty. Next, design of the three least square parameter estimators were explained that estimate the parameters of the control-oriented model. Then, AT-NMPC control problem was defined that uses two separate models: one for defining the objective function and another one for constraint handling. Using separate models ensures that constraints are handled based on the fixed nominal control-oriented model, while the objective function is defined based on an adapted control-oriented model to ensure that the optimal point of the cost function and the actual system are equivalent. The objective function of the controller was defined to minimize the energy cost while following a preceding vehicle.

	In the control evaluation step, simulations on high-fidelity model were performed by injecting uncertainties and delay into the simulation environment. Control evaluations showed that the proposed AT-NMPC is able to handle the defined constraints in the presence of uncertainty while improving the trip energy cost by 10\% compared to a tracking NMPC. Finally, an HIL experiment was conducted to show the real-time capability of the proposed controller, which showed that AT-NMPC has low computation costs while running on a prototype ECU.

	\section{Acknowledgment}
	
	Authors would like to gratefully express their appreciation to Toyota, NSERC and Ontario Centers of Excellence for financial support of this study.

	\bibliographystyle{IEEEtran}

	

	\end{document}